\documentclass[12pt,a4paper,twoside]{revtex4}  %{article}
\usepackage[francais]{babel}
\usepackage[T1]{fontenc}
\usepackage[dvips]{graphicx}
\usepackage{hangcaption} %pour des caption dont on contrôle la largeur
\usepackage{amsmath}
\usepackage{amsfonts}
\usepackage{euscript}
\usepackage{bbm} %pour écrire des lettres math creuses $\mathbbm{1}$ pour identité
\usepackage{enumerate}
\usepackage[english]{varioref} %pour des ref malines: page suivante, page précédente,...
\usepackage[section]{placeins}
\usepackage{subeqnarray}

\begin{document}

%NOTATIONS MATH TIREES DE THESE.STY
%partie relle
\renewcommand{\Re}{{\mathrm{\,Re\,}}}
%partie imag
\renewcommand{\Im}{{\mathrm{\,Im\,}}}
%Exp
\newcommand{\Exp}[1]{\textit{\large{e}}^{\,#1}}
%rect
\def\sq{\mathop{\rm rect}\nolimits}
%sinc
\def\sinc{\mathop{\rm sinc}\nolimits}
%erf
\def\erf{\mathop{\rm erf}\nolimits}
%sécante hyperbolique
\def\sech{\mathop{\rm sech}\nolimits}
%Fonction de Heaviside
\def\hea{\mathop{\Theta}\nolimits}
%Transformée de Fourier directe
\def\TF{\mathop{\raisebox{0.125em}{$\mathcal{F}$}}\nolimits}
%Transformée de Fourier inverse
\def\TFI{\mathop{\raisebox{0.125em}{$\mathcal{F}^{-1}$}}\nolimits}

\newcommand{\0}{^{(0)} }
\newcommand{\1}{^{(1)} }
\newcommand{\2}{^{(2)} }
\newcommand{\3}{^{(3)} }
\newcommand{\n}{^{(n)} }

%NOTATIONS TF ET CHAMP
%symbole complexe minuscule
\newcommand{\Cp}{\mathbb{C}}
%Complexe conjugué
\newcommand{\ccon}{^{\ast}} %dc E\ccon c'est le complexe conjugué de E
%Convolution
\newcommand{\conv}{{\ast}}
%Champ réel associé à E
\newcommand{\reE}{\mathcal{E}}
%Champ réel associé à n'importe quel champ
\newcommand{\re}[1]{\mathcal{#1}}
%Transformée de Fourier de n'importe quel champ
\newcommand{\tf}[1]{\tilde{#1}\/}
%Transformée de Fourier de E
\newcommand{\tfE}{\tf{E}}
%Transformée de Fourier de A
\newcommand{\tfA}{\tf{A}}

%Phase spectrale
\newcommand{\phiw}{\phi}
%Phase temporelle
\newcommand{\phit}{\varphi}

%%%%%%%%%%%%%%%%%%%%%%%%%%%%%%%%%%%%%%%%%%%%%%%%%%%%%%%%%%%%%%%%%%%%%%%%%%%%
%%%% CHAPITRE PULSE SHAPER %%%%%%%%%%%%%%%%%%%%%%%%%%%%%%%%%%%%%%%%%%%%%%%%%
%%%%%%%%%%%%%%%%%%%%%%%%%%%%%%%%%%%%%%%%%%%%%%%%%%%%%%%%%%%%%%%%%%%%%%%%%%%%

%MASQUE A CRISTAUX LIQUIDES
%largeur spectrale du masque à cl.
\newcommand{\dlM}{\Delta\lambda_{M}}
%largeur spatiale du masque
\newcommand{\lM}{L}
%nombre de pixels
\newcommand{\npix}{N}
%largeur spatiale d'un pixel+gap
\newcommand{\lpix}{l}
%largeur spectrale d'un pixel dispersion linéaire
\newcommand{\dopix}{\delta\omega}
%pulsation au pixel n, dispersion non-linéaire: \onl{n}
\newcommand{\onl}[1]{\omega_{#1}}
%réponse spectrale d'un pixel complete en approximation gaussienne \Rep(n,r,w)
\newcommand{\Rep}{A}
%Fenêtre temporelle du façonneur
\newcommand{\Trep}{T_0}
%épaisseur de la barrette (cl)
\newcommand{\epix}{{e_{CL}}}

%nom du masque
\newcommand{\M}{M}
%nom du masque appliqué à la position
\newcommand{\MX}{M_{X}}
%nom du masque appliqué à la pulsation
\newcommand{\Mw}{M_{w}}
%nom de la fonction de transfert totale dans l'approximation gaussienne
\newcommand{\Ht}{H}
%nom de la fonction de transfert idéale qu'on veut reproduire
\newcommand{\Hi}{H_i}
%nom de la fonction de transfert idéale qu'on veut reproduire au pixel n
\newcommand{\Hin}{H_{i,n}}
%nom de la fonction de transfert au niveau des gaps
\newcommand{\Hg}{G}
%nom de la fonction de transfert au niveau du gap n
\newcommand{\Hgn}{G_n}
%nom du ratio pixel/gap
\newcommand{\rat}{r}

%RESEAU
%angle d'incidence
\newcommand{\ti}{\theta_i}
%angle de diffraction
\newcommand{\td}{\theta_d}
%angle de diffraction de lambda centrale laser
\newcommand{\tdlc}{\theta_{d0}}
%séparation angulaire correspondant à  FWHM d'impulsion incidente
\newcommand{\Dtd}{\Delta\td}
%angle de Littrow
\newcommand{\tl}{\theta_l}

%LASER INCIDENT
%nom du champ
%\newcommand{\Ein}{E_{in}}
\newcommand{\Ein}{E_{E}}
%lambda centrale laser
\newcommand{\lc}{\lambda_0}
%omega centrale laser
\renewcommand{\oc}{\omega_0}
%largeur spectrale FWHM laser
\newcommand{\Dl}{\Delta\lambda_L}
%étalement spatial de largeur spectrale FWHM laser
\newcommand{\DLl}{\Delta L}
%largeur spectrale FWHM laser
\newcommand{\Do}{\Delta\omega_L}
%diamètre FWHM faisceau laser sur réseau
\newcommand{\Dx}{\Delta x}
%Durée FWHM intensité laser
\newcommand{\Dt}{\Delta t}

%LASER AU PLAN DE FOURIER
%nom du champ av masque
\newcommand{\Epfin}{E_{\textsf{pf},E}}
%nom du champ ap masque
\newcommand{\Epfout}{E_{\textsf{pf},S}}
%diamètre FWHM faisceau laser dans PF
\newcommand{\DX}{\Delta X}
%Pulsation dans le plan de Fourier correspondant à position X
\newcommand{\op}{\omega'}
%Largeur FWHM intensité spectral correspondant à diamètre laser dans PF
\newcommand{\Dop}{\Delta \omega'}
%Largeur FWHM intensité temporelle liée à convolution gaussienne ds PF
\newcommand{\DT}{\Delta T}

%LASER EN SORTIE
%nom du champ
%\newcommand{\Eout}{E_{out}}
\newcommand{\Eout}{E_{S}}
%Variable conjuguée de position dans plan, FWHM
\newcommand{\DK}{\Delta K}
%Couplage spatiotemporel: vitesse de couplage
\newcommand{\vc}{v}
%Impulsion chirpée: phase absolue
\newcommand{\tK}{\theta_K}

%CALIBRATION
%longueur d'onde de calibration
\newcommand{\lcal}{\lambda_{cal}}
%%%%%%%%%%%%%%%%%%%%%%%%%%%%%%%%%%%%%%%%%%%%%%%%%%%%%%%%%%%%%%%%%%%%%%%%%%%%

%%%%%%%%%%%%%%%%%%%%%%%%%%%%%%%%%%%%%%%%%%%%%%%%%%%%%%%%%%%%%%%%%%%%%%%%%%%%
%%%%%%%%%%%%% AUTRES SHAPERS %%%%%%%%%%%%%%%%%%%%%%%%%%%%%%%%%%%%%%%%%%%%%%%
%%%%%%%%%%%%%%%%%%%%%%%%%%%%%%%%%%%%%%%%%%%%%%%%%%%%%%%%%%%%%%%%%%%%%%%%%%%%
%Membrane de miroir def
\newcommand{\memb}{\mathcal{M}}
%Déphasage spectral dû à membrane de miroir def
\newcommand{\phimemb}{{\phi_\memb}}

%Champ entrée DAZZLER
\newcommand{\Edzi}{{E_i}}
%Champ diff DAZZLER
\newcommand{\Edzd}{{E_d}}
%Champ non-diff DAZZLER
\newcommand{\Edznd}{{E_{nd}}}
%Onde acoustique DAZZLER
\newcommand{\Adz}{{v}}
%Coef magique acousto-optique DAZZLER
\newcommand{\alpdz}{\alpha}
%Anisotropie optique DAZZLER
\newcommand{\Dndz}{{\Delta n}}
%Anisotropie optique de groupe DAZZLER
\newcommand{\Dngdz}{{\Dndz_{g}}}
%Vitesse acoustique DAZZLER
\newcommand{\vacdz}{{V}}
%Taille DAZZLER
\newcommand{\Ldz}{{L}}

%fréquence acoustique DAZZLER
\newcommand{\facdz}{{f_{ac}}}
%fréquence optique DAZZLER
\newcommand{\fopdz}{{f_{op}}}
%longueur d'onde optique DAZZLER
\newcommand{\lopdz}{{\lambda_{op}}}

%Fenêtre temporelle totale DAZZLER
\newcommand{\Tdz}{{T_{max}}}
%Délai de compensation DAZZLER
\newcommand{\Tcompdz}{{T_{cp}}}
%Fenêtre temporelle effective DAZZLER
\newcommand{\Teffdz}{{T_{eff}}}

%%%%%%%%%%%%%%%%%%%%%%%%%%%%%%%%%%%%%%%%%%%%%%%%%%%%%%%%%%%%%%%%%%%%%%%%%%%%

%%%%%%%%%%%%%%%%%%%%%%%%%%%%%%%%%%%%%%%%%%%%%%%%%%%%%%%%%%%%%%%%%%%%%%%%%%%%
%%%%%%%%%%% CHAPITRE TRANSITOIRES COHERENTS %%%%%%%%%%%%%%%%%%%%%%%%%%%%%%%%
%%%%%%%%%%%%%%%%%%%%%%%%%%%%%%%%%%%%%%%%%%%%%%%%%%%%%%%%%%%%%%%%%%%%%%%%%%%%

%%%%%Champ de pompe%%%%%
    %Champ de pompe temporel
    \newcommand{\Ep}{E_{pu}}
    %Champ de pompe réel
    \newcommand{\reEp}{\re{E}_{pu}}
    %Amplitude de pompe
    \newcommand{\Ap}{A_{pu}}
    %Phase temporelle de pompe
    \newcommand{\phitp}{\phit_{pu}}
    %Durée de pompe
    \newcommand{\dtp}{\Delta t_{pu}}
    \newcommand{\taup}{\tau_{pu}}
    %Durée de pompe sans chirp
    \newcommand{\dtpz}{{\dtp}_0} %z pour zéro
    %Coef de phase quadratique temporelle
    \newcommand{\alphap}{\alpha_{pu}}
    %Temps d'arrivée de Pulsation  résonnante avec transition g->e
    \newcommand{\teg}{t_{eg}}

    %Champ de pompe spectral
    \newcommand{\tfEp}{{\tf{E}_{pu}}}
    %Pulsation centrale de pompe
    \newcommand{\omp}{\omega_{pu}}
    %Longueur d'onde centrale de pompe
    \newcommand{\lamp}{\lambda_{pu}}
    %Phase spectrale de pompe
    \newcommand{\phiwp}{\phiw_{pu}}
    %Phase quadratique
    \newcommand{\phisecp}{{\phi\2_{pu}}}

    %Ecart à résonnance eg de Pulsation centrale de pompe wp-weg
    \newcommand{\domeg}{\delta \omega_{eg}}
%%%%%%%%%%%%%%%%%%%%%%%%

%%%%Séquence d'impulsion de pompe%%%%%

    %Mise en forme spectrale
    \newcommand{\hct}[1]{H^{#1}} % 1 est l'angle utilisé

    % champ séquence de pompe
    \newcommand{\EpSig}{{\Ep}_{\Sigma}}
    %1 pompe temporelle
    \newcommand{\Epu}{{\Ep}_{1}}
    %2 pompe temporelle
    \newcommand{\Epd}{{\Ep}_{2}}

    %1 pompe champ réel
    \newcommand{\reEpu}{{\reEp}_{1}}
    %2 pompe champ réel
    \newcommand{\reEpd}{{\reEp}_{2}}
    %1 Amplitude de pompe
    \newcommand{\Apu}{{\Ap}_{1}}
    %2 Amplitude de pompe
    \newcommand{\Apd}{{\Ap}_{2}}
    %1 Phase temporelle de pompe
    \newcommand{\phitpu}{{\phitp}_{1}}
    %2 Phase temporelle de pompe
    \newcommand{\phitpd}{{\phitp}_{2}}
    %1 Durée de pompe
    \newcommand{\dtpu}{{\dtp}_{1}}
    %2 Durée de pompe
    \newcommand{\dtpd}{{\dtp}_{2}}

    % champ spectral séquence de pompe
    \newcommand{\tfEpSig}{{\tfEp}_{\Sigma}}
    %1 Champ de pompe spectral
    \newcommand{\tfEpu}{{\tfEp}_{1}}
    %2 Champ de pompe spectral
    \newcommand{\tfEpd}{{\tfEp}_{2}}

    %1 Phase spectrale de pompe
    \newcommand{\phiwpu}{{\phiwp}_{1}}
    %2 Phase spectrale de pompe
    \newcommand{\phiwpd}{{\phiwp}_{2}}

%%%%%%%%%%%%%%%%%%%%%%%

%%%%%%Champ de sonde%%%%%
    %Champ de sonde temporel
    \newcommand{\Es}{E_{pr}}
    %Champ de sonde réel
    \newcommand{\reEs}{\re{E}_{pr}}
    %Amplitude de sonde
    \newcommand{\As}{A_{pr}}
    %Phase temporelle de sonde
    \newcommand{\phits}{\phit_{pr}}
    %Durée de sonde
    \newcommand{\dts}{\Delta t_{pr}}

    %Champ de sonde spectral
    \newcommand{\tfEs}{\tf{E}_{pr}}
    %Pulsation centrale de sonde
    \newcommand{\oms}{\omega_{pr}}
    %Longueur d'onde centrale de sonde
    \newcommand{\lams}{\lambda_s}
    %Phase spectrale de sonde
    \newcommand{\phiws}{\phiw_{pr}}

%Ecart à résonnance eg de Pulsation centrale de sonde ws-wfe
\newcommand{\domfe}{\delta \omega_{fe}}

%%%%%%%%%%%%%%%%%%%%%%%%%

%%%%%Délai pmpe-sde%%%%%
\newcommand{\tps}{\tau}
%%%%%%%%%%%%%%%%%%%%%%%%

%%%%%Syst 3 niveaux%%%%%
    %fonction d'onde atomique sur les trois niveaux:
    \newcommand{\Psia}[1]{\left| \Psi{#1} \right\rangle}
    %niveau fondamental
    \newcommand{\nivg}{\left| g \right\rangle}
    %amplitude de probabilité niveau fondamental
    \newcommand{\apg}{a_g}
    %Population niveau fondamental
    \newcommand{\Pg}{P_{\!g}}
    %Pulsation de niveau g
    \newcommand{\omg}{\omega_{g}}

    %niveau excité
    \newcommand{\nive}{\left| e \right\rangle}
    % durée de vie niveau excité
    \newcommand{\taue}{\tau_{e}}
    %amplitude de probabilité niveau excité
    \newcommand{\ape}{a_e}
    %amplitude de probabilité niveau excité séquence de ppe
    \newcommand{\apeSig}{{a_e}_{\Sigma}}
    %amplitude de probabilité niveau excité ppe1
    \newcommand{\apeu}{{a_e}_1}
    %amplitude de probabilité niveau excité
    \newcommand{\aped}{{a_e}_2}
    %Reconstruction géométrique: amplitude de probabilité niveau excité +
    \newcommand{\apedp}{{a^+_e}_{\!\!2}}
    %Reconstruction géométrique: amplitude de probabilité niveau excité -
    \newcommand{\apedm}{{a^-_e}_{\!\!2}}
    %Population niveau excité
    \newcommand{\Pe}{P_{\!e}}
    %Pulsation de niveau e
    \newcommand{\ome}{\omega_{e}}

    %niveau i
    \newcommand{\nivi}{\left| i \right\rangle}
    % durée de vie niveau excité
    \newcommand{\taui}{\tau_{i}}
    %amplitude de probabilité niveau excité
    \newcommand{\api}{a_i}
    %amplitude de probabilité niveau excité séquence de ppe
    \newcommand{\apiSig}{{a_i}_{\Sigma}}
    %amplitude de probabilité niveau excité ppe1
    \newcommand{\apiu}{{a_i}_1}
    %amplitude de probabilité niveau excité
    \newcommand{\apid}{{a_i}_2}
    %Reconstruction géométrique: amplitude de probabilité niveau excité +
    \newcommand{\apidp}{{a^+_i}_{\!\!2}}
    %Reconstruction géométrique: amplitude de probabilité niveau excité -
    \newcommand{\apidm}{{a^-_i}_{\!\!2}}
    %Pulsation de niveau i
    \newcommand{\omi}{\omega_{i}}

    %niveau final
    % durée de vie niveau final
    \newcommand{\tauf}{\tau_{f}}
    \newcommand{\nivf}{\left| f \right\rangle}
    %amplitude de probabilité niveau final
    \newcommand{\apf}{{a_f}}
    %amplitude de probabilité avec facteur de phase
    \newcommand{\bpf}{{b_f}} %\apf(\infty,\tps)=\Exp{i\omfe\tps}\bpf(\tps)

    %amplitude de probabilité avec facteur de phase ppe1
    \newcommand{\bpfu}{{\bpf}_{1}} %\apf(\infty,\tps)=\Exp{i\omfe\tps}\bpf(\tps)
    %amplitude de probabilité avec facteur de phase ppe2
    \newcommand{\bpfd}{{\bpf}_{2}} %\apf(\infty,\tps)=\Exp{i\omfe\tps}\bpf(\tps)

    %amplitude de probabilité niveau final ppe1
    \newcommand{\apfu}{{\apf}_{1}}
    %amplitude de probabilité final ppe2
    \newcommand{\apfd}{{\apf}_{2}}

    %Population niveau final
    \newcommand{\Pf}{{P_{\!f}}}
    %Pulsation de niveau f
    \newcommand{\omf}{\omega_{f}}

    %Pulsation de transition g->e
    \newcommand{\omeg}{\omega_{eg}}
    %Longueur d'onde de transition g->e
    \newcommand{\lambdaeg}{\lambda_{eg}}
    %Largeur Doppler de transition g->e
    \newcommand{\doppeg}{\Delta_{eg}}
    %Pulsation de transition e->f
    \newcommand{\omfe}{{\omega_{fe}}}
    %Dipole de transition g->e
    \newcommand{\mueg}{\mu_{eg}}
    %Dipole  de transition e->f
    \newcommand{\mufe}{\mu_{fe}}
    %Largeur Doppler de transition e->f
    \newcommand{\doppfe}{\Delta_{fe}}

 %Pulsation de transition g->i
    \newcommand{\omig}{\omega_{ig}}

%%%%%%%%%%%%%%%%%%%%%%%%

%%%%Plusieurs états finaux%%%%

    %niveaux finaux
    \newcommand{\nivn}{\left| n \right\rangle}
    %amplitude de probabilité niveau final
    \newcommand{\apn}{{a_n}}
    %amplitude de probabilité avec facteur de phase
    \newcommand{\bpn}{{b_n}} %\apn(\infty,\tps)=\Exp{i\omne\tps}\bpn(\tps)

    %amplitude de probabilité niveau final avec facteur de phase ppe1
    \newcommand{\bpnu}{{\bpn}_{1}}
    %amplitude de probabilité final avec facteur de phase ppe2
    \newcommand{\bpnd}{{\bpn}_{2}}

    %amplitude de probabilité niveau final ppe1
    \newcommand{\apnu}{{\apn}_{1}}
    %amplitude de probabilité final ppe2
    \newcommand{\apnd}{{\apn}_{2}}

    %Population niveau final
    \newcommand{\Pn}{{P_n}}
    %Pulsation de niveau n
    \newcommand{\omn}{\omega_{n}}
    %Pulsation de transition e->n
    \newcommand{\omne}{{\omega_{ne}}}
    %Dipole  de transition e->n
    \newcommand{\mune}{\mu_{ne}}

    %Fonction de réponse convoluant la pompe
        %pour donner signal mesuré
    \newcommand{\h}{h}
    %Fonction de transfert correspondant
    \newcommand{\tfh}{\tf{\h}}
%%%%%%%%%%%%%%%%%%%%%%%%%%%%%%

%%%%%%Signal de fluo %%%%%%
    %signal de fluo pour un angle donné:
    \newcommand{\fluo}[1]{S^{\,#1}} %\fluo{} pour S, \fluo{\theta} pour S^theta

%%%Signal reconstruit%%%%
    \newcommand{\Gg}{G}
 %signal reconstruit est dérivée de af(tau)
    %signal temporel
    \newcommand{\g}{g}
    %signal spectral
    \newcommand{\tfg}{\tf{\g}}
    %phase spectrale
    \newcommand{\phiwg}{\phiw_{g}}
%%Durée de pompe
%\newcommand{\taum}{\tau_m}
%%Pulsation centrale de pompe
%\newcommand{\omm}{\omega_m}
%%%%%%%%%%%%%%%%%%%%%%%%

%%Notations Math%%%%%%%%
%Partie Principale de Cauchy
\newcommand{\Pp}{\mathcal{P_P}}
%%%%%%%%%%%%%%%%%%%%%%%%

%%%%%%Silberberg-Jerome%%%%%%%%
%Phase de saut de phase de pi
\newcommand{\phistep}{{\phi_{\pi}}}
%Pulsation saut de phase de pi
\newcommand{\omstep}{{\omega_{\pi}}}
%longueur d'onde saut de phase de pi
\newcommand{\lambdastep}{{\lambda_{\pi}}}
%Temps d'arrivée saut de phase de pi
\newcommand{\tstep}{{t_{\pi}}}
%%%%%%%%%%%%%%%%%%%%%%%%
%%%%%%%%%%%%%%%%%%%%%%%%%%%%%%%%%%%%%%%%%%%%%%%%%%%%%%%%%%%%%%%%%%%%%%%%%%%%

\title{Real time Quantum state holography using coherent transients}

\author{Antoine Monmayrant}
%\address
\affiliation{Laboratoire Collisions, Agr\'egats, R\'eactivit\'e (UMR
5589 CNRS-UPS), IRSAMC, Universit\'e Paul Sabatier Toulouse 3, 31062
Toulouse cedex 9, France}
\author{B\'eatrice Chatel}
\affiliation{Laboratoire Collisions, Agr\'egats, R\'eactivit\'e (UMR
5589 CNRS-UPS), IRSAMC, Universit\'e Paul Sabatier Toulouse 3, 31062
Toulouse cedex 9, France}
\author{Bertrand Girard}
\affiliation{Laboratoire Collisions, Agr\'egats, R\'eactivit\'e (UMR
5589 CNRS-UPS), IRSAMC, Universit\'e Paul Sabatier Toulouse 3, 31062
Toulouse cedex 9, France}
%{Laboratoire de Collisions, Agr\'egats et R\'eactivit\'e (CNRS UMR 5589), IRSAMC\\ Universit\'e Paul Sabatier,\\  31062 Toulouse CEDEX, France\\
%E-mail:beatrice@irsamc.ups-tlse.fr}

\date{\today}

\begin{abstract}
In a two level atom, real-time quantum state holography is performed
through interferences between quantum states created by a reference
pulse and a chirped pulse resulting in coherent transients. A
sequence of several measurements allows one to measure the real and
imaginary parts of the excited state wave function. These
measurements are performed during the interaction with the
ultrashort laser pulse. The extreme sensitivity of this method to
the pulse shape provides a tool for electric field measurement.
\end{abstract}

%\ocis{(320.5540) Pulse shaping; (320.7100) Ultrafast measurements}

\maketitle

\section{Introduction}
The effect of laser pulse shape on a quantum system is related to
the nature of the interaction. For a linear response of the system
(one-photon transition in the weak field regime), the final outcome
depends only on the spectral component at the resonance frequency
and is therefore independent of the pulse shape, and particularly of
the spectral phase \cite{Bouchene98KCs}. This explains for instance
why signals equivalent to wave-packet interferences could be
observed with incoherent light as well as with ultrashort pulses
\cite{Jones95IncoherentLight}. However, the temporal evolution
towards the final state may depend strongly on the pulse shape. A
straightforward illustration of this statement is the non-resonant
interaction which leads to transient excitation of the system, but
to no final excitation. In the absence of predesigned control
mechanisms
 only a closed loop scheme \cite{Judson92,Warren93} may be employed to find efficient pulse shapes \cite{Assion98,Levis01,Motzkus02bio,woste-co-sience2003}:
The outcome of many different shapes is fed back into an algorithm
that iteratively optimizes the excitation shape without insight into
the physical mechanism that is triggered by a particular shape.

In contrast the effect of shapes on small systems can be
systematically studied within an open-loop scheme
\cite{silberberg98,Leone02WPphasecontrol,silberberg02,Degert02CTshaped}.
This open-loop approach is well adapted to these systems for which
theoretical predictions are reliable. It consists of reaching a
specific goal (manipulation of the temporal response of a system
excited by a light pulse) without any experimental feed-back.
Physical analysis of the process allows one to predetermine the
theoretical pulse shape which leads to the desired result. It is
then implemented experimentally.

In this article, we describe manipulation of Coherent Transients
(CT) in an open loop approach. These CT are oscillations in the
excited state population resulting from the interaction between a
two-level system and a weak chirped pulse. The shape of these
oscillations is extremely sensitive to slight changes in the pulse
shape \cite{RbShapingAPB04,Degert02CTshaped}. Two previous letters
\cite{MonmayrantCT-reconstruction-05,MonmayrantCT-spirograph-05}
have shown that their high sensitivity provides a new scheme for
quantum state measurement and electric field reconstruction. This
article presents in details the works and calculations corresponding
to these results. First we recall the coherent transients and how to
manipulate them. Then the quantum state measurement reconstruction
is presented in detail. In particular several schemes are discussed.
Then the experimental set-up and several previous feasibility test
are described. Finally the results are presented and discussed.

\section{Coherent Transients principle}

\begin{figure} [ht]
\centering
\includegraphics[width=4cm]{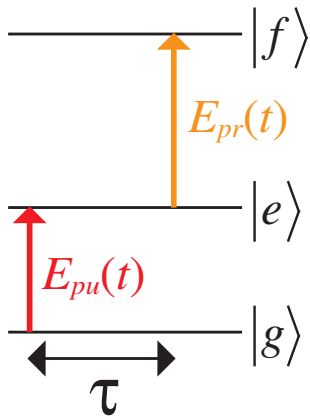}
 \caption[levels]{Excitation scheme.} \label{levels}
\end{figure}

\begin{figure} [ht]
\centering
\includegraphics[width=8cm]{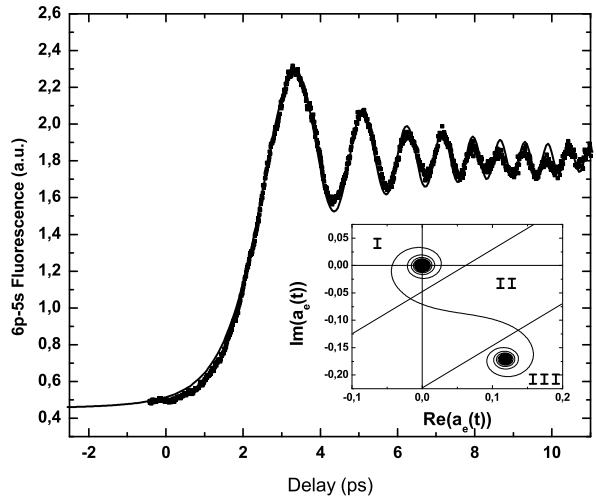}
\caption[NormalCT]{Experimental Coherent Transients on Rb
(5s-5p$_{1/2}$ at $\lambda = 795 \, {\rm nm}$), for a chirp of $-8.
\, 10^5 \, {\rm fs}^2$ (dots) and the corresponding simulation
obtained by numerical resolution of the Schr\"{o}dinger equation
(solid line) \cite{zamith01}. Inset : Theoretical excited state
amplitude drawn in the complex plane. } \label{NormalCT}
\end{figure}

The CT result from the interaction of a two-level system ($\nivg$
and $\nive$) with a chirped pulse $\Ep(t)$ of carrier angular
frequency $\omp$ close to resonance ($\omp\simeq\omeg$). The
transient excited state population is probed towards the $\nivf$
level in real time by a second ultrashort pulse $\Es(t)$ which is
Fourier transform limited and short compared to the characteristic
features of $\Ep(t)$. Its frequency is close to resonance ($\omfe$).
The fluorescence arising from the $\nivf$ state is then recorded as
a function of the pump-probe delay $\tau$ (cf Fig. \ref{levels}).
The probe pulse provides access to the temporal evolution of the
population in $\nive$, produced by the pump beam. The result of the
interaction is described by first order perturbation theory, and the
fluorescence is proportional to
\begin{eqnarray}
S(\tau)&=&|a_f(\tau)|^2\nonumber \\ &\propto&
\left|\int_{-\infty}^{+\infty} \Es(t-\tau)\exp(i\omfe(t-\tau))
a_e(t)dt \right|^2
\end{eqnarray}
%\mu_{eg}\mu_{fe}/4\hbar^2
 with
\begin{eqnarray}\label{a_e}
a_e(t)&=& \int_{-\infty}^t \Ep(t')\exp(i\omeg t')dt' \nonumber \\
&=& \int_{-\infty}^t \Exp {-(t'/\taup)^2} \Exp {-i (\domeg t'+
\alphap t'^2)}dt'
\end{eqnarray}
in the case of a simply chirped pulse $\Ep(t)= \Exp {-(t/\taup)^2}
\Exp {-i (\omp t+ \alphap t^2)}$. Here $\domeg = \omp - \omeg$ is
the resonance mismatch, $\taup$ the pulse duration and $\alphap$ the
chirp rate. A quadratic phase appears in the integral giving
$a_e(t)$ (Eq. \ref{a_e}), leading to oscillations of the probability
$|a_f(\tau)|^2$ as already demonstrated
\cite{zamith01,Degert02CTshaped} (cf Fig. \ref{NormalCT}). These
strong oscillations result from interferences between the
probability amplitude excited at resonance and after resonance.

The CT phenomenon is better understood by examining the behavior of
$a_e \left( t \right)$ in the complex plane as displayed in the
inset of Fig. \ref{NormalCT}. The probability amplitude follows a
Cornu spiral starting from the origin. Three regions can be
distinguished. The two spirals result from contributions before (I)
and after (III) resonance for which the quadratic phase varies
rapidly. The intermediate region (II) corresponds to the passage
through resonance where the phase is stationary. It provides the
main contribution to the population. The two spirals, although
similar, result in totally different behaviors of the population.
The first one (I) winds round the origin with an increasing radius.
The resulting probability increases thus slowly and regularly and
remains small. After resonance (III), a second spiral winds round
the asymptotic value leading to strong oscillations of the
population.

We show in the next section how a modification of the excitation
scheme provides the possibility to observe oscillations due to the
first part of the pulse.

\section{Quantum state measurements}
\subsection{principle}

The CTs are extremely sensitive to tiny phase modifications of the
pump pulse \cite{Degert02CTshaped,RbShapingAPB04}. Therefore, they
can provide detailed information on the exciting pulse and
simultaneously on the excited quantum state. However,
%the present scheme cannot achieve a complete pulse measurement since it is based on a probability measurement. , one should overcome the
although sensitive to phase effects these CTs give access to the
excited state probability $|a_e(\tau)|^2$ whereas the probability
amplitude is necessary to achieve a complete measurement of the
electric field. Moreover, the oscillations are only produced by the
second part of the pulse (after resonance)\cite{zamith01}. To
overcome these limitations, we propose a new excitation scheme based
on a two pulse sequence with a well defined phase relationship. The
pump pulse is written as
\begin{equation}
\Ep(t)=\Epu(t)+e^{i\theta}\Epd(t)
\end{equation}
where $\Epu(t)$ and $\Epd(t)$ are two replica of the initial pulse
%generated by splitting the same initial pulse $\Ep(t)$ and adding
with additional spectral phase. These can be obtained with either a
Michelson-type interferometer or a pulse shaper. The first pulse
$\Epu(t)$ creates an initial population in the excited state. The
second pulse $\Epd(t)$ is strongly chirped and sufficiently delayed
so that it does not overlap with the first pulse. This second pulse
creates a probability amplitude in the excited state which
interferes with the initial probability amplitude created by the
first pulse.

It should be noted that the details of the shape of the first pulse
are not critical. Only the final state reached at the end of the
first pulse is involved in the temporal evolution of the system
during the second pulse.
\begin{figure} [ht]
\centering
\includegraphics[width=0.6\textwidth]{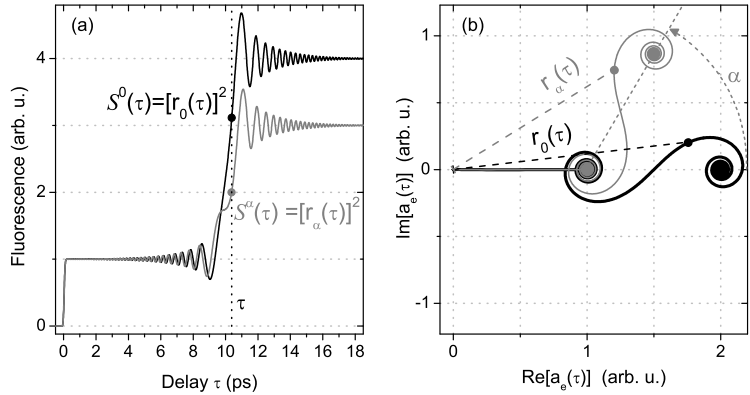}
\caption{ (a) Theoretical CTs scans for a geometric reconstruction:
$\theta=0$ (black), $\theta=\alpha$ (gray), with $\phiwp\1= 10 \;
{\rm ps}$, $\phiwp\2=2.10^5 {\rm \;fs^2}$. (b) Corresponding
probability amplitudes (same color code).}\label{s2-ct-cornu}
\end{figure}

Looking at the evolution of the quantum state in the complex plane
(Fig. \ref{s2-ct-cornu}~(b)), one sees that the effect of the first
pulse is to shift the starting point of the spiral so that
oscillations due to CTs occur on the whole duration of the second
pulse. Assuming a sufficient time interval between the two pulses to
avoid any overlap, the probability amplitude induced by the first
pulse $\apeu(t)$ has reached its asymptotic value $\apeu(\infty)$
when the interaction with the second pulse starts. For a probe pulse
significantly shorter than the details one wants to retrieve on the
excited state population, the recorded fluorescence
$\fluo{\theta}(\tps)$ is directly proportional to the excited state
population. During (or after) the second pulse, it can be written as
\begin{eqnarray}\label{cc:eq:cttheta-final}
\nonumber\fluo{\theta}(\tps)&=&\left|\apeu(\infty)+\Exp{i\theta}\aped(\tps)\right|^2\\
&=&\left|\apeu(\infty)\right|^2+\left|\aped(\tps)\right|^2+2\Re\left[\Exp{i\theta}\apeu\ccon(\infty)\aped(\tps)\right]
\end{eqnarray}
$\left|\apeu(\infty)\right|^2$ can be deduced from a measurement of
$\fluo{\theta}(\tps)$ in the interval between the two exciting
pulses. In order to determine the complex number associated to the
probability amplitude, at least a second measurement is necessary as
described in the next subsection.

\subsection{Reconstruction techniques}
\label{reconstruction} The probability amplitude produced by the
second pulse $\Epd(t)$ is retrieved by combining the CTs scans
$\fluo{\theta}(\tps)$ (see Eq. \ref{cc:eq:cttheta-final}) obtained
for different values of the programmable phase $\theta$.
% (phase cycling).
The goal here is to extract the cross term
$\apeu\ccon(\infty)\aped(\tps)$ from a set of scans. The factor
$\apeu\ccon(\infty)$ is deduced - except for its phase - from the
fluorescence observed at the end of the first pulse.
%In the following, we might omit this factor for the sake of simplicity.
 We will show here different possible
reconstruction schemes. As an example, we simulate the CTs
corresponding to the following two-pulse sequence: the first pulse
is 100~fs long, the second one is chirped to 10~ps ($2.10^5$~fs$^2$
quadratic phase) and delayed by 6 or 10~ps. Both pulses are resonant
$\omp=\omeg$ and have the same energy. Determining the real and
imaginary part of the probability amplitude requires at least two
equations, which means two CT scans with different values of
$\theta$. In this case, a system of two second order equations is
obtained. A geometric method is used to solve it. With a third
measurement, the quadratic term in Eq. \ref{cc:eq:cttheta-final} can
be removed in order to obtain a linear system.

\begin{figure}[ht]
\centering
\includegraphics[width=0.6\textwidth]{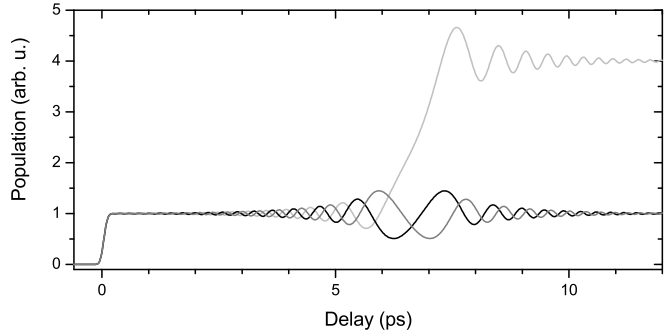}
\caption{Theoretical CTs for a three-scans reconstruction: pulse
sequence with $\phiwp\1= 6 \; {\rm ps}$, $\phiwp\2=-2.10^5 {\rm
\;fs^2}$ and $\theta_k=2k\pi/3$. $\theta_0=0$  : light grey line;
$\theta_1=2\pi/3$ : black line; $\theta_2=4\pi/3$ : grey line.}
\label{s3-ct}
\end{figure}

With a set of three scans, the angles $\theta_k (k=0,2)$ must be
chosen so that the corresponding matrix is not singular. This is the
case for instance with  $\theta_k=2k\pi/3$. The corresponding CTs
are plotted in Fig. \ref{s3-ct}.
From these measurements, we calculate $\fluo{[3]}(\tps)$ defined as:
\begin{equation}\label{cc:eq:fluo-rec-3bis}
\fluo{[3]}(\tps)=  \frac{1}{3} \fluo{0}(\tps) - \frac{(1+i\sqrt
{3})}{6} \fluo{2\pi/3}(\tps) -\frac{(1-i\sqrt {3})}{6}
\fluo{4\pi/3}(\tps)= \apeu\ccon(\infty)\aped(\tps)
\end{equation}
%Similar results can also be obtained by replacing the scan with  $\theta=\pi$ by a scan with the second pulse $\Epd(t)$ only, providing $\fluo{\emptyset}(t)$. In this case, one gets
%$\fluo{[3]}(\tps)$:
%\begin{equation}\label{cc:eq:fluo-rec-3}  \fluo{[3']}(\tps)=  \frac{\fluo{0}(\tps)-i\fluo{\pi/2}(\tps) -(1-i)\fluo{\emptyset}(\tps)}{2} = \apeu\ccon(\infty)\aped(\tps)+\frac{1-i}{2}\left|\apeu(\infty)\right|^2 \end{equation}
%in which $\left|\apeu(\infty)\right|^2$ can be measured by looking at the plateau  on the CTs for times within the pulse separation (between 1 and 2~ps in figure \ref{s3-ct}).

Alternatively, the probability amplitude can be retrieved from a set
of two CT measurements provided that a system of two nonlinear
equations is solved. For two different values of $\theta$, for
example 0 and $\alpha\neq 0, \; \pi$, we thus have to solve the
two-equation system:
\begin{subeqnarray}\label{cc:eq:s2-ct-syst}
\slabel{cc:eq:s2-ct-eq1}
\fluo{0}(\tps)&=&\left|\apeu(\infty)\right|^2+\left|\aped(\tps)\right|^2+2\Re\left[\apeu\ccon(\infty)\aped(\tps)\right]\\
\slabel{cc:eq:s2-ct-eq2}
\fluo{\alpha}(\tps)&=&\left|\apeu(\infty)\right|^2+\left|\aped(\tps)\right|^2+2\Re\left[\Exp{i\alpha}\apeu\ccon(\infty)\aped(\tps)\right]
\end{subeqnarray}
If the second pulse is much weaker than the first one, the quadratic
term in $\left|\aped(\tps)\right|$ can be neglected to obtain a
simple linear equation system. In this case one easily obtains a
unique solution and $\alpha =\pi/2$ is the simplest choice.

Generally, the non-linear equation system gives two different
solutions and only one is physically acceptable. To easily identify
this solution and separate it from the other one, we have developed
a geometric reconstruction which is described in detail in the
appendix.

\section {Experiment} \label{Experiment}
\subsection{Experimental set-up}
\label{setup}

\begin{figure}[!ht]
\begin{center}
\includegraphics
[width=0.45\textwidth] {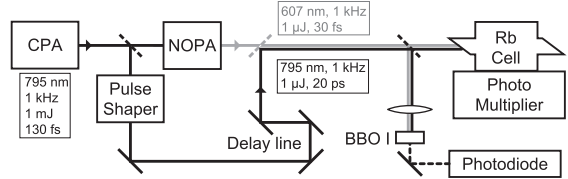} \caption {Experimental set-up.
NOPA : non colinear optical parametric amplifier; CPA : chirped
pulse amplifier} \label{Fig-setup}
\end{center}
\end{figure}
The experimental set-up is displayed in Fig. \ref{Fig-setup}. The 5s
- 5p (P$_{1/2}$) transition (at 795 nm) is resonantly excited with a
pulse sequence. The transient excited state population is probed "in
real time" on the (5p - (8s, 6d)) transitions with an ultrashort
pulse (at 607 nm). The laser system is based on a conventional Ti:
Sapphire laser with chirped pulse amplification (Spitfire Spectra
Physics) which supplies 1 mJ -130 fs -795 nm pulses. Half of the
beam is used for the pump pulse. The remaining seeds a home made
Non-collinear Optical Parametric Amplifier (NOPA) compressed using
double pass silica prisms, which delivers pulses of a few
microJoule, 30 fs -FWHM pulse intensity, centered around 607 nm. The
pump pulse is shaped with a programmable pulse-shaping device
producing the pulse sequence, recombined with the probe pulse and
sent into a sealed rubidium cell. The pump-probe signal is detected
by monitoring the fluorescence at 420 nm due to the radiative
cascade (8s,~6d)~-~6p~-~5s collected by a photomultiplier tube as a
function of the pump-probe delay. In parallel, a cross-correlation
of the pump pulse sequence is recorded. The pulse shaping device is
a 4f set-up composed of one pair each of reflective gratings and
cylindrical mirrors. Its active elements -two 640 pixels liquid
crystal masks- are installed in the common focal plane of both
mirrors. This provides high resolution pulse shaping in phase and
amplitude \cite{pulseshaperRSI04}. This is used to generate the
shaped pump pulse sequence by applying the function
\begin{equation}
\hct{\theta}(\omega)=\frac{1}{2}\mathbbm{1}+%
\frac{1}{2}\exp\left[i\theta+i\phiwp\1(\omega-\omp)+i\frac{\phisecp}{2}(\omega-\omp)^2\right]
\end{equation} The laser is centered at resonance
($\omp=\omeg$).

\subsection{Interferometric stability}
\label{stability}
 The relative stability of the two pulse sequence
is a crucial point in the present experiment. Both the relative
phase and delay between the two pulses of the pump sequence should
be kept stable as compared to $2\pi$ or to the optical period $T_o$.
Experiments of wave packet interferences have been performed with a
Michelson interferometer used to produce the pulse pair. The delay
was either actively \cite{Scherer90,Scherer91,belabas-shaping-ir-01}
or passively
\cite{Jones93RamseyPRL,Blanch95,Blanch97Cs,Blanch98Cs2,Ohmori03}
stabilized using different techniques. The best achieved stability
is better than $T_o/100$ with a Michelson placed under vacuum
\cite{Ohmori03}.

Alternatively, in experiments where only the amplitude of the
interference pattern is needed, different strategies have been
developed. These are based either on periodic modulation of the
delay followed by a lock-in amplifier
\cite{Broers93,Charalambidis_PRL_atto}, or random fluctuations
applied to the delay followed by measurement of the resulting noise
\cite{Averbukh_COIN95PRL,Shapiro_COIN_JCP98,Bucksbaum_QSholography_PRL98}.

\begin{figure}[ht]
\begin{center}
\includegraphics [width=0.45\textwidth]
{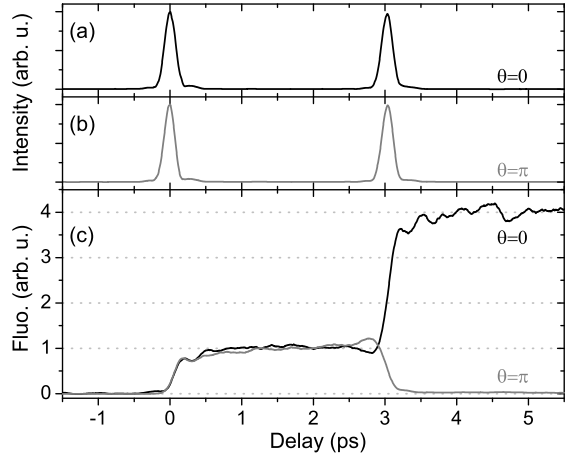} \caption {Experiments with a pump sequence of two
Fourier Limited pulses centered at 795 nm and a 25 fs probe pulse at
607 nm, as a function of the probe pulse delay. Pump-probe
cross-correlations for a relative phase of $0$ (a) or $\pi$ (b). (c)
Fluorescence from the 8s-6d states for the two relative phases
(black line: $\theta=0$; Gray line : $\theta=\pi$).}
\label{2pulse_delay}
\end{center}
\end{figure}

In our approach, the required stability and control of the phase and
delay are naturally provided by the phase and amplitude pulse shaper
\cite{pulseshaperRSI04}. As an illustration, we have performed
demonstration experiments with a pump sequence consisting of two
identical Fourier transform limited pulses, delayed by 3 ps. In a
first example (see Fig. \ref{2pulse_delay}), the relative phase (at
the resonance frequency) is set to 0 and $\pi$ for two scans of the
pump - probe delay. Two cross-correlations measurements (Fig.
\ref{2pulse_delay}a and b) illustrate the relative position of the
pulses. The phase shift of $\pi$ does not affect these
cross-correlations. In the pump-probe scan, the three positions of
the probe pulse with respect to the pump sequence lead to: (i) No
fluorescence signal when the probe is before the pump pulses, (ii) a
constant signal independent of the relative phase for the probe
before the pump pulses, (iii) constructive ($\theta=0$) or
destructive ($\theta=\pi$) interferences for a probe pulse after the
pump sequence. In the constructive interference case, the
fluorescence signal is 4 times the signal obtained with a single
pulse, as expected from usual interferences.

\begin{figure}[!ht]
\begin{center}
\includegraphics
[width=0.45\textwidth] {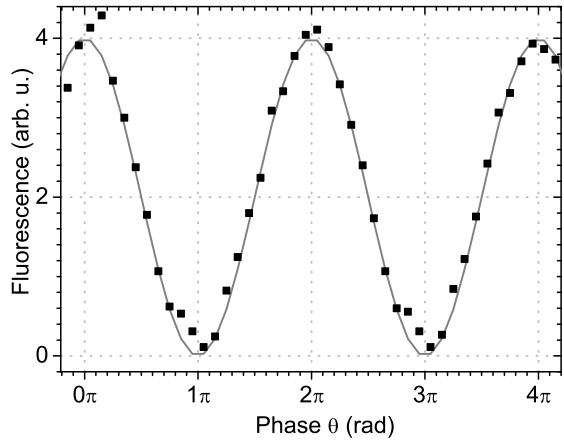} \caption {Same scheme
as in Fig. \ref{2pulse_delay} but with a fixed pump - probe delay
and a variable relative phase. Experiment (Squares) and sine fit
(solid grey line).} \label{2pulse_phase}
\end{center}
\end{figure}

In a second experiment, the pump-control delay is set to a constant
value of $267 \, {\rm ps}$ and the relative phase is scanned (Fig.
\ref{2pulse_phase}). These two results illustrate both the excellent
stability of the set-up and the control over the programmable phase.

\section{Results and discussion}
%\subsection{Quantum state measurement}
We present a series of results obtained with a sequence of two
pulses generated by the high resolution phase and amplitude pulse
shaper: The first one is close to Fourier limited (a replica of the
input pulse) and the second one is time delayed ($\phiwp\1= 6 \;
{\rm ps}$) and strongly chirped ($\phiwp\2=-2.10^5 {\rm \;fs^2}$).
Their amplitude are set equal. With phase and amplitude shaping, an
extra relative phase $\theta$ can easily be added to the second
pulse. The various records correspond to different values of
$\theta$ and are used to illustrate the two reconstruction methods
described in Section \ref{reconstruction}.

\begin{figure}[htb]
\centering
\includegraphics[width=9cm]{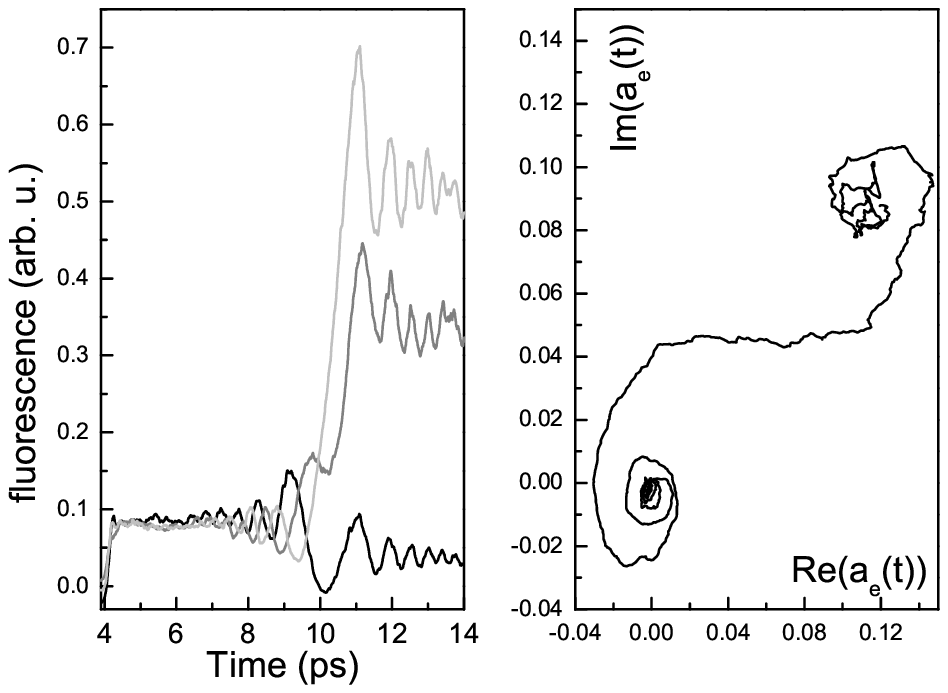
} \caption{a) Experimental Coherent Transients resulting from the
excitation of the atom by a FT limited pulse (at time $\tau=0$)
followed by a chirped pulse ($\phiwp\1= 6 \; {\rm ps}$,
$\phiwp\2=-2.10^5 {\rm \;fs^2}$), for three different relative
phases $\theta_0 \simeq 0.7$ (light grey line), $\theta = \theta_0 +
2\pi/3$ (black line) and $\theta = \theta_0 +4\pi/3$ (dark grey
line) between the two pulses. b) Probability amplitude
$a_{e,2}(\tau)$ reconstructed from the three measurements presented
in a), using a linear reconstruction and displayed in the complex
plane. The Cornu spiral appears clearly.}\label{FTresultlin}
\end{figure}

In the first example, three scans with phases separated by $2\pi/3$
are used: $\theta = \theta_0 + 2 k \pi/3 \; (k=0,1,2)$. The quality
of the reconstruction does not depend on $\theta_0$ and here we have
$\theta_0 \simeq 0.7$ (the reconstruction efficiency does not depend
on $\theta_0$). The scans are displayed in Fig. \ref{FTresultlin}-a.
As a difference to the case of a single chirped pulse (Fig.
\ref{NormalCT}) \cite{zamith01}, the three regimes are now clearly
visible. The oscillations are observed before resonance as well as
after resonance. The behavior during the passage through resonance
depends directly on the relative phase $\theta$. A rapid increase,
slow increase or slow decrease is observed resulting from
constructive, partially constructive or destructive interferences.
As expected, and similarly to the case of two FT limited pulses (see
Fig. \ref{2pulse_delay} and \ref{2pulse_phase}), the asymptotic
value depends also strongly on $\theta$. The linear reconstruction
method is used. The good stabilities of the laser and experimental
set-up allow us using directly the raw data without any adjustment.
The excited state probability amplitude produced by the second pulse
(with $\theta = \theta_0$) is extracted from the three measurements
and displayed in Fig.\ref{FTresultlin}-b). One observes clearly the
expected Cornu spiral.

\begin{figure}[htb]
\centering
\includegraphics[width=9cm]{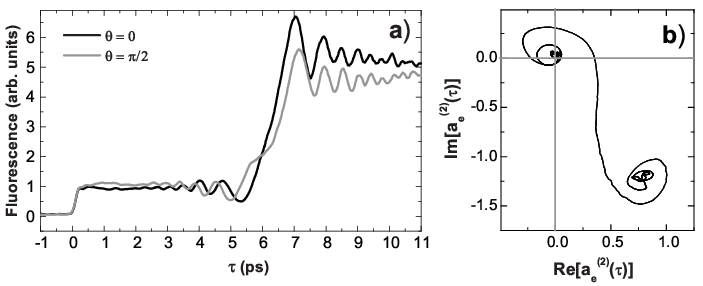
} \caption{a) Experimental Coherent Transients resulting from the
excitation of the atom by a FT limited pulse (at time $\tau=0$)
followed by a chirped pulse ($\phiwp\1= 6 \; {\rm ps}$,
$\phiwp\2=-2.10^5 {\rm \;fs^2}$), for two different relative phases
$\theta_0 \simeq -0.8$, $\theta = \theta_0 + \pi/2$ between the two
pulses. b) Probability amplitude $a_{e,2}(\tau)$ reconstructed from
the two measurements presented in a) and displayed in the complex
plane. The Cornu spiral appears clearly.}\label{FTresult}
\end{figure}

\begin{figure}[htb]
\centering
\includegraphics[width=7cm]{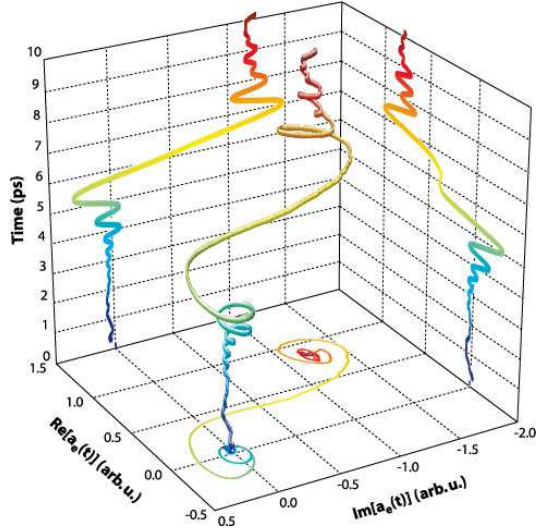}
\caption{3D spiral representing the time evolution of the excited
state probability amplitude (same data as in Fig. \ref{FTresult}).
The vertical axis represents the time.}\label{spirale}
\end{figure}

In the second example displayed in Fig. \ref{FTresult}, two scans
with phases separated by $\pi/2$: $\theta = \theta_0$ and $\theta_0
+ \pi/2$ are used for the nonlinear reconstruction. Here $\theta_0
\simeq -0.8$. The nonlinear method requires determining separately
the population $\left|\apeu(\infty)\right|^2$ created by the first
pulse. The plateau immediately after the end of the first pulse is
used for this purpose. The excited state probability amplitude
produced by the second pulse and extracted from the two measurements
is displayed in Fig. \ref{FTresult}b. The reconstructed probability
amplitude is also displayed in Fig. \ref{spirale} in a 3D plot (real
and imaginary part of the probability amplitude as a function of
time). The projections on the various 2D planes are also displayed.
The expected Cornu spiral \cite{zamith01} is clearly seen in the
complex plane.

In previous experiments \cite{zamith01,Degert02CTshaped}, only the
excited state probability was measured. Here, the initial
preparation of a coherent superposition of $\nive$ and $\nivg$ by
the first pulse allowed measuring the probability amplitude in real
time during its evolution in interaction with the laser pulse.

The two methods provide similar quality of reconstruction. The
linear approach requires three measurements. It increases the
recording time by a factor of 1.5 as compared to the nonlinear
method. Conversely it is more robust and can be used in a wider
variety of situations (relative phase, intensity ...). A larger
number of recordings could be combined in a linear square fit
approach to improve the accuracy of the measurement. This would of
course be at the extent of the recording time.

Several examples of quantum phase measurements of states created by
ultrashort pulses are based on interferences between an unknown wave
function and a ''reference'' wave function. These wave functions are
created by a sequence of two ultrashort pulses (an unknown pulse and
a reference pulse). The quantum state created by the unknown pulse
is deduced either by time- and frequency- integrated fluorescence
measured as a function of the delay \cite{SchleichShapiro98PRL}, or
by measuring the population of each eigenstate for different values
of the relative phases \cite{Yeazell_holography_97}. Alternatively,
the amplitude of noise resulting from random fluctuations of the
delay is measured
\cite{Bucksbaum_QSholography_PRL98,Bucksbaum_holography_Nature99}.
In another approach, the dispersed fluorescence emitted by an
oscillating nuclear wave packet in a diatomic molecule was recorded
as a function of time \cite{Walmsley95}. In this case, the
fluorescence wavelength - position relationship is derived from the
Franck-Condon principle.

In all these examples involving several excited states, either a
particular selectivity is used to detect independently each excited
state, or the delay is used to obtain a set of measurements which
are then inverted to obtain the amplitude of each quantum state. In
our study, only one single excited state is involved and the
measurements are performed at the same delay. This ensures
determining the temporal evolution of the quantum state.

Our quantum state measurement method can be extended to the case of
$p$ excited states $(\nivi)_{i=1,p}$ of different energies $\hbar
\omi$. Their probability amplitudes can be retrieved from a set of
$2p+1$ measurements in a linear reconstruction scheme. As an
example, the first measurement can be performed with the second
pulse $\Epd(t)$ only, providing
\begin{equation}
\fluo{\emptyset}(\tps)= \sum\limits_{i = 1}^p {\left| {\apid(\tps)}
\right|^2}
\end{equation}
This allows thus to remove the nonlinear contributions from the
subsequent measurements. The remaining $2p$ measurements are
performed with the two pulse sequence, each with a set of $p$ phases
$\left( \theta_{i,k} \right)_{i=1,p}$ for $k=1, 2p$ applied at the
frequencies $\omi$. They provide with the quantity
\begin{equation}\label{cttheta-pstates}
\fluo{}_k(\tps) =\sum\limits_{i = 1}^p
{\left|\apiu(\infty)\right|^2}+\sum\limits_{i = 1}^p
{\left|\apid(\tps)\right|^2}+2\sum\limits_{i = 1}^p
{\Re\left[\Exp{i\theta_{i,k}}\apiu\ccon(\infty)\apid(\tps)\right]}
\end{equation}
As in the case with a single excited pulse, $\sum\limits_{i = 1}^p
{\left|\apiu(\infty)\right|^2}$ can be deduced from a measurement of
$\fluo{}_k(\tps)$ in the interval between the two exciting pulses.
Since the phases can be chosen independently, it is always possible
to find a set of phases for which the system of $2p$ linear
equations can be inverted. This would not be the case if the phases
were not applied independently but through an extra delay $\tps '_k$
(giving $\theta_{i,k} = \omig \tps '_k$).

\section{Conclusion}

We have presented a new method to determine the real time evolution
of an excited quantum state in interaction with an ultrashort laser
pulse.

By simple derivation of the excited state probability amplitude, it
is possible to retrieve the electric field (phase and amplitude) of
the second pump pulse $\Epd(t)$, provided that the probe pulse is
well known or short enough. The possibilities offered by this
technique are discussed in detail elsewhere
\cite{MonmayrantCT-reconstruction-05}. It can also be used for a
differential measurement to analyze the changes induced by inserting
a material. In this last case, the requirements on the properties of
the probe pulse are less severe.

We thank Chris Meier for fruitful discussions.
\section{Appendix}

We detail here the geometrical reconstruction used to solve the set
of second-order non-linear equations. This latter interprets the
equation system (\ref{cc:eq:s2-ct-syst}) in terms of circles
intersection in the complex plane. Figure \ref{s2-ct-cornu}~(a)
shows the two CTs scans for $\theta=0$ (black) and $\theta=\alpha$
(gray) used for the reconstruction (in the simulations
$\alpha=\pi/3$). The corresponding probability amplitudes are
plotted in Fig. \ref{s2-ct-cornu}~(b). In both cases, the
contribution of the first Fourier-limited pulse is a straight line
and the contribution of the second pulse is a Cornu spiral. The
phase $\theta$ only changes the relative orientation of the line and
the spiral. At any time $\tau$, the CTs values $\fluo{0}(\tps)$ and
$\fluo{\alpha}(\tps)$ respectively correspond to $r_{0}^2(\tps)$ and
$r_{\alpha}^2(\tps)$, where $r_{0}(\tps)$ and $r_{\alpha}(\tps)$ are
the distances in the complex plane between the origin and the
current positions on both spirals (see Fig. \ref{s2-ct-cornu}~(b)).
Retrieving the probability amplitude produced by the second pulse
corresponds to geometrically reconstructing the black Cornu Spiral
in Fig. \ref{s2-ct-cornu}~(b), using the two time dependant
distances $r_{0}(\tps)$ and $r_{\alpha}(\tps)$.

\begin{figure} [ht]
\centering
\includegraphics[width=0.4\textwidth]{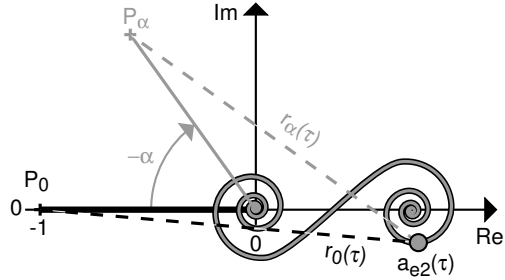}
\caption{New geometric interpretation: $r_{0}(\tps)$ and
 $r_{\alpha}(\tps)$ are the distances between $\aped(\tps)$ and two reference points $P_0$ et $P_\alpha$.}%
 \label{rec-cercle2-nb}
\end{figure}

\begin{figure} [ht]
\centering
\includegraphics[width=0.4\textwidth]{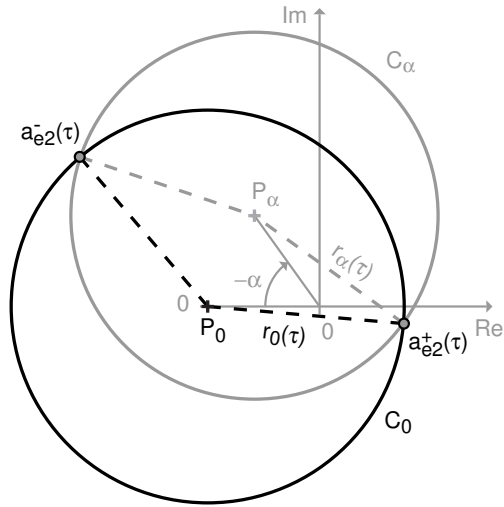}
\caption{Geometric reconstruction of $\aped(\tps)$. Two solutions
$\apedp(\tps)$
and $\apedm(\tps)$ are available; The physical one starts in (0,0).}%
\label{rec-cercle3-nb}
\end{figure}

To achieve this, we mentally rotate the gray path by an angle of
$-\alpha$, around the starting point of the Cornu Spiral (1,0). We
then choose this point as the new origin for the complex plane.
These transformations preserve both angles and distances and
therefore do not change our equation system. Figure
\ref{rec-cercle2-nb} shows the two paths after the transformations.
We call $P_0$ and $P_\alpha$ the starting points of each path whose
coordinates are known: $(-1,0)$ and $(-\cos(\alpha),\sin(\alpha))$
respectively. Thanks to these transformations, the two Cornu Spirals
are now superimposed and correspond to the amplitude probability we
want to retrieve. The two distances $r_{0}(\tps)$ and
$r_{\alpha}(\tps)$ can now be seen as the distances between the
point $\aped(\tps)$ on the Cornu spiral and two reference points
$P_0$ and $P_\alpha$.
To geometrically reconstruct $\aped(\tps)$ one just needs to find,
for every time $\tps$, the intersection of the circle
$\mathcal{C}_0$ (centered on $P_0$ with a radius $r_0(\tps)$) and
the circle $\mathcal{C}_\alpha$ (centered on $P_\alpha$ with a
radius $r_\alpha(\tps)$), as depicted in figure
\ref{rec-cercle3-nb}.
We get two different solutions, $\apedp(\tps)$ and $\apedm(\tps)$.
The physically acceptable one starts in (0,0). To avoid degeneracy,
the Cornu spiral should not cross the $(P_{0}, P_{\alpha})$ line.
Two ways of pushing the spiral away from $(P_{0}, P_{\alpha})$ are
increasing the intensity of the first pulse, and reducing the angle
$\alpha$. However, a too small angle leads to near-degeneracy of the
circles, increasing thus the uncertainties in determining their
crossing points. Usually, the reconstruction works well with a first
pulse at least as intense as the second one and an angle $\alpha$ in
the interval $[\pi/4,\pi/2]$.

%LA BIBLIO
%\nocite{*}  % pour faire apparaître tous les ouvrages
\bibliography{bib}
\bibliographystyle{falseosajnl}
%\listoffigures
%%%%%%%%
\end{document}